\documentclass{rstransa}
\begin{document}
\title{Gravity Beyond Einstein? Yes, but in Which Direction?}
\author{Demosthenes Kazanas$^1$, Demetrios Papadopoulos$^2$ \& Dimitris Christodoulou$^3$}
\address{$^1$NASA/Goddard Space Flight Center, Code 663, Greenbelt, MD 20771 $^2$Aristotelian University, Thessaloniki, Greece, $^3$ University of Massachusetts Lowell, Lowell, MA 01854}
\subject{gravitational theory, dark matter...}
\corres{demos.kazanas@nasa.gov}

\begin{abstract}
We present qualitative arguments in favor of an extension of the theory of the gravitational interaction beyond that resulting from the Hilbert-Einstein action. To this end we consider a locally conformal invariant theory of gravity, discussed some thirty years ago by Mannheim and Kazanas. We discuss its exact solution of the static, spherically symmetric configurations and, based on these, we revisit some of the outstanding problems associated with gravity, high energy interactions and sketch potential resolutions within the conformal gravity framework.
\end{abstract}

\begin{fmtext}

\end{fmtext}

\maketitle
....
\section{Introduction}
The introduction of General Relativity (hereafter GR) and its confirmation by the classic tests of light bending and the account of the precession of Mercury, is  without any doubt one of the great intellectual achievements. Since that time GR matured to a subject of its own with great in-depth mathematical analysis and numerous observations and applications, including situations where the background space-time deviates significantly from that of Minkowski (e.g. black holes, neutron stars) or very little from that, but where high accuracy is demanded (e.g GPS).  Numerous tests of various aspects of GR have been performed with results consistent with its basic premises, typically summed-up in the non-scientific press with the expression ``Einstein proven right again!".

Thus, it is generally considered that the only significant development left in the field of gravitational physics is that of the quantization of GR, i.e. the development of a procedure that will allow consideration of quantum gravitational effects in a way done for example in QED. However, these are considered to be present at energy/mass scales close to that of Plank $M_P \simeq 10^{19}$ GeV, far removed from those of the laboratory or for that any observable in the present universe (with the exception of CMB polarization).

In the way of such an achievement stands the dimensionful coupling constant of gravity, namely Newton's constant, $G$, which precludes an approach emulating that of QED or similar approaches of other interactions, whose Lagrangians involve only dimensionless constants (apparently superstrings are a way of achieving that but it will not be considered in the present exposition). Furthermore, the presence of a god-given dimensionful constant in the gravitational Lagrangian leads to the presence of black holes (hereafter BH) within the theory, constructs that imply the presence of and presumably hide singularities, situations that the theory fails to treat, which presumably will be cured by a quantum version of the theory.

It is generally considered that, in the absence of a quantum theory of gravity, the issue of singularities can be at presently tacitly ignored, alleviated by their hiding behind horizons, surfaces of infinite redshift of which no information can escape. However, this notion brought up a different issue with the discovery by Hawking that the presence of horizons leads to the evaporation of the black holes; the horizons effectively disrupt the quantum correlations of the vacuum fluctuations allowed by the Uncertainty Principle, leading to the appearance of half of the quanta (e.g. photons, $e^+e^-$-pairs) of these fluctuations on one side of the horizon (Hawking radiation), while keeping the other half (of negative energy) on the horizon interior side to reduce the BH mass, which slowly evaporates (the existence of such radiation is generic to the presence of any horizon, not only to those of black holes; the radiation registered on one side of the horizon of e.g.  an accelerating detector called Unruh radiation). While the effect of Hawking radiation is well formulated and it is not considered at present controversial, nonetheless, it leads to the so-called `information loss paradox', namely to the fact that the information of Hawking radiation is all random, contrary to that which formed the BH; this implies violation of unitarity, one of the cherished principles of Quantum Mechanics. To the best assessment of the authors of this note a well accepted resolution of this paradox is currently lacking.

Besides the above issues, there are several others involving the gravitational interaction, but these are generally attributed to our poor knowledge of the matter content and distribution that produces the gravitational field which leads to the observed kinematics, rather than to deficiencies of our understanding of the gravitational interaction.  These are the so-called Dark Matter and Dark Energy issues.

The first one concerns the rotation curves of spiral galaxies, which do not exhibit the expected $v_{\phi} \sim r^{-1/2}$ Keplerian fall-off at galactocentric distances large  compared to that of their exponentially decaying surface brightness (the rotational velocities remain constant with increasing distance or ``flat" as they are commonly referred to). This kinematic behavior is broadly attributed to the presence of haloes of non-radiant, non-baryonic Dark Matter (DM) which is thought to surround the disks of spiral galaxies (and also contribute to the dynamics of ellipticals, which exhibit velocity dispersions in disagreement with the stellar matter content implied by their light distribution). It is generally assumed that this component consists of some sort of unknown yet particle associated with the high energy physics interactions (supersymmetric particles were the most favored candidates, but the most recent LHC experiments failed discover any).  One should also note that even in the presence of such particles, "flat" rotation curves demand rather fine tuning between the amount and distributions of luminous and dark matter.

The second issue concerns the apparent lack of deceleration in the expansion of the universe, in fact an acceleration, at late times. This was determined by measuring the flux of a large number of type Ia supernovae, considered standard candles, that occur at a specific redshifts $z$ and forming the ensuing redshift-magnitude relation. Fits of the evolution of the universe to this relation is best achieved by adding a cosmological term, $\Lambda/3$, to the equations that describe the scale factor evolution of an homogeneous and isotropic universe\cite{1}. Its energy density is such that it contributes roughly 70\% to the present closure density of the universe, a value some 120 orders of magnitude smaller than the "natural" value of such a term, which it is believed it ought to be of order of the Planck energy density.

A rather interesting fact, noted some time ago\cite{2}, is that the introduction of DM to account for the galactic rotation curves becomes necessary at regions where the acceleration of the orbiting stars, $a$, drops  below a characteristic value  $a_0 \sim 10^{-8}$ cm s$^{-2}$; interestingly the value of this acceleration is close to $c H_0$ ($c$ is the speed of light and $H_0$ the value of the Hubble constant), suggesting a cosmological origin of the underlying dynamics. A similar fact,  not as much noted in the literature, is that the introduction of Dark Energy as a means of accounting for the acceleration of the universe has occurred when the acceleration of its expansion $\ddot R(t)$ dropped below this same value $a_0$.

To return to the essential aspects of the gravitational theory, one should note that GR (i.e. Einstein gravity), despite its impressive achievements, is the only theory among the fundamental interactions that is formulated not on a local invariance (gauge) principle but only on general covariance and the demand that the resulting equations be second order; thus, it provides, in short, a covariant formulation of the Newtonian potential and the Poisson equation that relates the former to its matter source. This, then, raises the question of whether a covariant formulation of gravity based on a local invariance principle is necessary or even desirable. Besides the broader underlying theoretical considerations, such an enterprise should be considered only if it can provide some (not necessarily complete) resolution of the outstanding issues associated with Einstein gravity, some of which have been outlined above.

\section{Conformal Gravity}

Motivated by considerations such as those of the last paragraph, Mannheim \& Kazanas \cite{3}, decided to look at a theory of gravity that incorporates such a local invariance principle. As such they have chosen that of scale invariance, i.e. a theory whose Lagrangian remains invariant under local stretchings of the geometry of the form $g_{\mu \nu} \rightarrow \Omega(x)^2 g_{\mu \nu}$.

As noted by Weyl, one can introduce a local invariance in any theory by enlarging the derivative operator, in the case of conformal invariance by defining the gauge invariant derivative
\begin{equation}\label{CovDeriv}
D_{\mu} = \partial_{\mu} + \kappa_{\mu} ~~{\rm with}~~ \kappa_{\mu}=\frac{\Omega_{\mu}}{\Omega}~~{\rm being ~the ~Weyl~ vector}
\end{equation}

Under this definition of the derivative the Christoffel symbols become
\begin{equation}\label{Chris}
 \Gamma^{\lambda}{\mu \nu} \rightarrow  \Gamma^{\lambda}{\mu \nu} +\frac{1}{\Omega}(\delta^{\lambda}_{\mu}\Omega_{\nu} + \delta^{\lambda}_{\nu}\Omega_{\mu} - g_{\mu \nu} \Omega^{\lambda})
\end{equation}
and then, with a redefinition of the covariant derivative that employs the new Christoffel symbols, obtain the gauge invariant Riemann, Ricci tensors and Ricci scalar\footnote{The interested reader can find a very lucid description of this analysis in \cite{4}}, which by necessity involve also the vector $\kappa_{\mu}$. There is however a combination of $R^{\lambda}_{\mu \rho \nu}, R_{\mu \nu}$ and $R^{\mu}_{\mu}$, namely the Weyl tensor, $C^{\lambda}_{\mu \rho \nu}$ which is \textit{independent} of the vector $\kappa_{\mu}$; as a result, the action
\begin{equation}\label{action}
  I_W = -\alpha \int d^4x (-g)^{1/2}
C_{\lambda\mu\nu\kappa}C^{\lambda\mu \nu\kappa}
=-2\alpha \int d^4x (-g)^{1/2}(R_{\lambda\mu}R^{\lambda
\mu}-(R^{\alpha}_{\phantom{\alpha} \alpha})^2 /3) + {\rm total~derivative}
\end{equation}
is the unique, conformally invariant action obtained without the introduction of additional fields and with $\alpha$ a purely dimensionless constant (the Gauss-Bonnet total derivative has been employed to eliminate the square of the Riemann tensor from the $C_{\lambda\mu\nu\kappa}C^{\lambda\mu \nu\kappa}$ expression to obtain the RHS of Eq. (2.3).

%This led by Milgrom\cite{2} to the introduction of Modified Newtonian Dynamics (in short MOND) \cite{1}. This theory proposes to replace Newton's law $F = ma$ by $F = m\sqrt{a \, a_0}$ when $a \ll a_0$, with a smooth transition between the two regimes. Interestingly the acceleration $a_0 \simeq c \, H_0$ appears to have cosmological significance ($c$ is the speed of light and $H_0$ the Hubble constant).

\subsection{Exact Solutions}
In the following we provide some exact solutions of the conformal theory with references to the original works.

%\smallskip
%\noindent \textit{1. The Static Spherically Symmetric Solution}
\subsubsection{The Static Spherically Symmetric Solution}

The price to be paid for enlarging the symmetries of the gravitational action is the higher order of the resulting equations (generally of 4th order).  Nonetheless, \cite{3} exploiting the conformal invariance of the theory they were able to obtain an exact solution for the static, spherically symmetric problem of the theory (the equivalent of the Schwarzschild solution of GR), which reads
\begin{equation}\label{sss}
  -g_{00}= 1/g_{rr}= B(r)=1 - 3 \beta \gamma - \beta(2-3 \beta \gamma )/r
+ \gamma r - kr^2
\end{equation}
where $\beta, \gamma,$ and $k$ are three appropriate dimensionful integration constants. 
%This solution contains the familiar exterior Schwarzschild solution (thereby yielding the desired exterior Newtonian potential term) together with two additional potential terms. The $kr^2$ term represents a background de Sitter geometry. 
This solution contains the familiar exterior Schwarzschild solution (thereby yielding the desired exterior Newtonian potential term) together with two additional potential terms: The $kr^2$ term is well understood and represents a background de Sitter geometry (a vacuum solution in the present theory). Finally, the quark confining-type linear $\gamma r$ term is a new gravitational term and is a special feature of the fourth order theory, in particular the conformal one\footnote{We note in passing that the vacuum, static, spherically symmetric solution of the $(R^{\mu}_{\mu})^2$ Lagrangian, while it produces the $\beta/r$ and $k r^2$ terms, it does not yield the linear $\gamma r$ term.}. This term is associated with conformal degrees of freedom, as the solution with $\beta = 0$ is shown to be conformally flat\cite{3}\footnote{The expressions of the gravitational field equations for the general spherically symmetric problem are given in \cite{5}; their most general expressions appear too complicated to be useful, but simplified, specific cases may provide novel insights to the interested reader.}.

\begin{figure}[!h]
\centering\includegraphics[width=2.5in]{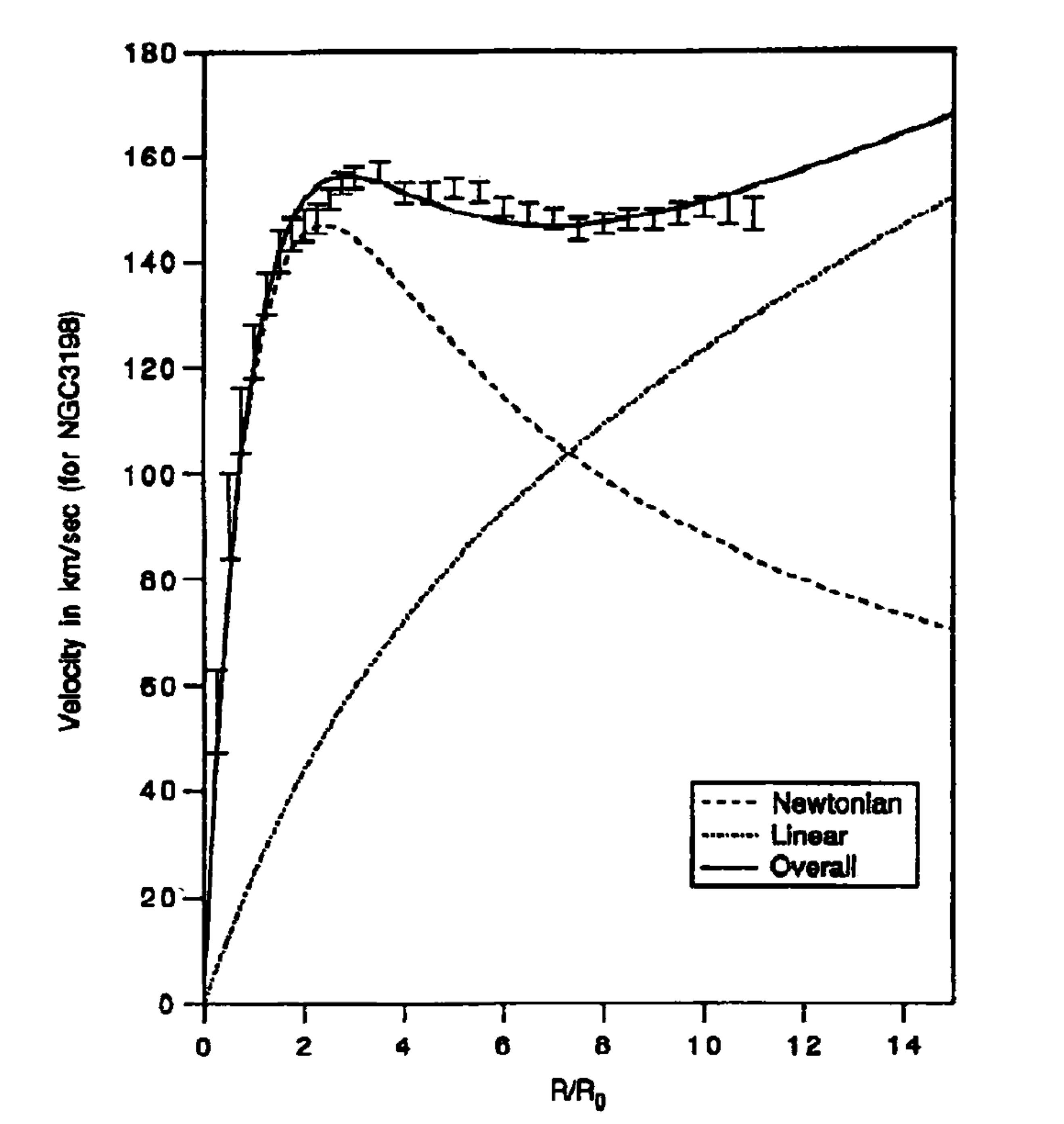}
%%% where xxxxxx name represents "figurename.eps"
\caption{The HI galactic rotation velocity curve data of NGC 3198, as a function of the galactocentric distance, in units of its surface brightness exponential length $R_0$  ($\Sigma = \Sigma_0 e^{-R/R_0}, R_0 = 2.72 {\rm kpc}$ (from \cite{6}) along with the associated Newtonian contribution (dashed line), which peaks at $R=2.2 R_0$, and the Weyl gravity linear term $\gamma r$ contribution (dotted line)(from \cite{7}).  }
\label{fig1}
\end{figure}

It was discussed in \cite{3}, that the apparent asymptotic non-flatness of the $\gamma r, k r^2$ terms, implies their magnitudes to be of order $R_H$ and $R_H^2$, with $R_H$ the Hubble radius (see however \cite{7} for a different point of view). Thus for sufficiently small (stellar) values of $r$, the conformal theory would appear to enjoy the experimental successes of static Einstein gravity, which have been obtained on solar system or smaller distance scales; however, for the assumed value of $\gamma$, on galactic scales, the Newtonian and the linear terms appeared to be of the same order of magnitude providing a potential resolution of the issue of galactic rotation curves without the need to introduce Dark Matter. An example of such an approach is shown in figure \ref{fig1}, where the contributions of the Newtonian, the linear terms and their total contribution to the rotation curves are shown along with the HI rotation curve data of the galaxy NGC 3198.

Taking the derivative of the $g_{00}$ term with respect to $r$ (and ignoring the much smaller contribution of the $k r^2$ term on galactic scales) one obtains an effective force per unit mass (acceleration)
\begin{equation}\label{acceleration}
  a = -\frac{\beta(2-3\beta\gamma}{r^2} - \gamma)
\end{equation}
One should note that with the assumed scalings, the value of $\gamma$ is of order $c H_0$, as proposed in \cite{2} in order to resolve the "flat" galactic rotation curves without introducing Dark Matter. To our view, it is significant that an acceleration of the proper magnitude occurs in the solution of a theory without any \textit{a priori} interest in the specific problem. We believe this not to be coincidental, even though the precise details of applications of this effect on the astrophysical setting and its ramifications remain at present rather obscure. It is not at present, clear even whether the characteristic relevant quantity that forces deviations from Newtonian dynamics, is an acceleration or a column density
\begin{equation}\label{NH}
  \frac{M}{r^2} \simeq \frac{c^2}{G R_H}\simeq 1 {\rm g ~cm}^{-2} \simeq 6 \times 10^{23}~  {\rm particles~ cm}^{-2}
\end{equation}
obtained on replacing $\beta$ by $GM/c^2$, the gravitational radius. One should note though, that, contrary to GR, this theory "does not know" about Newton's constant $G$, which in this instance we have borrowed from our notions of Newtonian theory. It is also of interest to note that this value of the column, is consistent with those of galaxies and close to that of the universe.

%\smallskip
%\noindent \textit{2. The Static Spherically Symmetric Electrovac Solution}
\subsubsection{The Static Spherically Symmetric Electrovac Solution}

The conformal symmetry imposed on the gravitational side of the field equations, implies that non-vacuum solutions must also involve, on the $T_{\mu \nu}$ matter side of the equations, also a conformally invariant stress energy momentum tensor. An obvious such tensor is that of a single particle of charge $Q$; the equivalent solutions are generally called $electrovac$ solutions, the Reissner - Nordstr\"om one being the equivalent one of GR\footnote{The exact static spherically symmetric electrovac and stationary axisymmetric and stationary axisymmetric electrovac solutions are given in \cite{8}.}.

As shown in \cite{3} the $W^r_r$ component of the gravitational tensor is given by %
\begin{eqnarray}
 \nonumber % Remove numbering (before each equation)
  W^r_r & = & W^{rr}/B(r) = B^{\prime}B^{\prime \prime \prime}/6 -
B^{\prime \prime 2}/12 - (B B^{\prime \prime \prime}- B^{\prime}B^{\prime \prime})/3r \\
   & & - (B B^{\prime \prime} + B^{\prime 2})/3r^2
+2 B B^{\prime}/3r^3- B^2/3 r^4+ 1/3 r^4
\end{eqnarray}\label{wrr}
%
%
%\begin{equation}\label{w11}
% W^r_r = B^{\prime}B^{\prime \prime \prime}/6 -
%B^{\prime \prime 2}/12 - (B B^{\prime \prime \prime}- B^{\prime}B^{\prime \prime})/3r \\
%- (B B^{\prime \prime} + B^{\prime 2})/3r^2
%+2 B B^{\prime}/3r^3- B^2/3 r^4+ 1/3 r^4
%\end{equation}
while the corresponding component of the EM tensor of a point charge $Q$ is given by
\begin{equation}\label{Trr}
  T^r_r = \frac{Q^2}{2 r^4}
\end{equation}
On setting $W^r_r = T^r_r$ and following the procedure given in \cite{3} one obtains
\begin{equation}\label{qrr}
 -g_{00}= 1/g_{rr} = B(r)=1 - 3 \beta \gamma  - \beta(2-3 \beta \gamma )/r
+ \gamma r - Q^2/8 \alpha \gamma r - kr^2
\end{equation}
One should note immediately the difference between this solution and the equivalent one of GR (the Reissner - Nordstr\"om solution): The contribution of the charge $Q$ to the geometry has the same dependence as that of the mass, namely a $1/r$ potential. This is due to the dimensionless coupling constant of the theory, $\alpha$, compared to Newton's constant $G$ of GR, which has dimension $[r^2]$. In the conformal case, to make this component of the metric dimensionless, the parameter $\gamma$ appears in the denominator of this term. As a result, there is no regime in $r$ for which the charge term would dominate that of the mass, as in the GR solution with the $G Q^2/r^2$ charge contribution to the metric, an issue that thought to present an "in principle" problem at sufficiently small values of $r$.

%\smallskip
%\noindent \textit{3. Newtonian Limit}
\subsubsection{Newtonian Limit}

As noted earlier, Einstein gravity, with insistence in second degree equations, provides in its static solutions a covariant generalization of Poisson's equation. This raises immediately the question on where does conformal gravity stands on this issue. With a more detailed discussion given in \cite{7}, we present here a brief summary.

Besides the explicit expression of $W^r_r$ given in \cite{3} and in Eq. (2.7), one can easily obtain the value of $W^{00}$ from \cite{5}. This turns out to be
\begin{eqnarray}
 \nonumber % Remove numbering (before each equation)
  W^{00} &=& -B^{\prime \prime \prime \prime}/3 + {B^{\prime \prime}}^2/12B
-B^{\prime \prime \prime} B^{\prime}/6B - B^{\prime \prime \prime}/r
-B^{\prime \prime} B^{\prime}/3rB  \\
   & & +B^{\prime \prime}/3r^2 + {B^{\prime}}^2/3r^2B - 2 B^{\prime}/3r^3
-1/3r^4B + B/3r^4 .
\end{eqnarray}
Despite the non-linearity of each of Eqs. (2.7, 2.10) the combination
\begin{equation}\label{00-rr}
  3(W^0_{{\phantom 0} 0} - W^r_{{\phantom r} r})/B =
B^{\prime \prime \prime \prime} + 4 B^{\prime \prime \prime}/r =
(rB)^{\prime \prime \prime \prime}/r= \nabla^4B
\end{equation}
is a linear, fourth order operator on $B$. One should note that this is an exact rather than perturbative relation within conformal gravity, even though a similar form could possibly appear in some linearized version of other fourth order theories. One should further note that in Einstein gravity no combination of the Einstein tensor $G^{\mu \nu}$ components yields the Laplacian as an exact expression, but only as perturbative linearization. Then, in the static spherically symmetric situation one obtains for the metric coefficient $B$ the expression
\begin{equation}\label{del4}
\nabla^4 B(r) = f(r) = {3 (T^0_{{\phantom 0} 0} - T^r_{{\phantom r} r})/4
\alpha B(r)}
\end{equation}
with $\alpha$ the dimensionless coupling constant of the $C^2$ Lagrangian. Considering that the Green's function of the operator $\nabla^4$ is $\vert
{\bf r} - {\bf r^{\prime}} \vert$ one can obtain $B(r)$ by integration over a source  $f(r)$
\begin{equation}\label{B}
  B(r) =  - {1 \over 8 \pi} \int d^3{\bf r^{\prime}} f(r^{\prime})\vert
{\bf r} - {\bf r^{\prime}} \vert
= - {1 \over 12 r} \int dr^{\prime} f(r^{\prime}) r^{\prime}
[\vert r^{\prime}+r \vert ^3 - \vert r^{\prime} - r
\vert^3]
\end{equation}
obtained after performing the angular integrations. Exterior to the source the solution then takes the form
\begin{equation}\label{r>R}
  B(r>R) =- {1 \over 6} \int_{0}^{R} dr^{\prime} f(r^{\prime})
[3{r^{\prime}}^2 r+{r^{\prime}}^4 /r]
\end{equation}
with the interior the solution is given by
\begin{equation}\label{r<R}
  B(r<R) =
-{1 \over 6} \int_{0}^{r} dr^{\prime} f(r^{\prime})
[3{r^{\prime}}^2 r +{r^{\prime}}^4 /r]
-{1 \over 6} \int_{r}^{R} dr^{\prime} f(r^{\prime})
[3{r^{\prime}}^3+{r^2 r^{\prime}}]
\end{equation}

One thus sees that the Newtonian and linear terms are interior moments of the source distribution $f(r)$
\begin{equation}\label{bg}
  \beta(2-3\beta \gamma)
= {1 \over 6} \int_{0}^{R} dr^{\prime} f(r^{\prime}) {r^{\prime}}^4
~~~,~~~\gamma = -{1 \over 2} \int_{0}^{R} dr^{\prime} f(r^{\prime})
{r^{\prime}}^2
\end{equation}
while the second term of Eq. (2.15) indicates that interior to a spherically symmetric shell, the space is generally de Sitter rather than Minkowski as is in GR. We are thus able to obtain all terms of the full solution as moments of the (conformal) matter distribution.

While the expressions of Eq. (2.16) indicate that the Newtonian and the linear potentials can be obtained as interior moments of the source, it is apparent that $f(r)$ cannot be a delta function, since this would yield zero for its 4th moment. In conformal theory therefore, sources cannot be point-like, not a surprising conclusion since they have to be stretchable. The natural potential between point sources in the theory is the linear term, with the Newtonian $1/r$ term obtainable as the left-over of integration of the linear term over \textit{extended sources}, making the Newtonian potential the \textit{short distance limit} of the theory, in complete disagreement to most treatments that either attempt to treat gravity in analogy with EM (given the $1/r$ potentials of both) or consider 4th order Lagrangians in addition with that of GR and its dimensionful coupling constant $G$.

\subsection{General Considerations}
%\smallskip
%\textit{1. Conformal Gravity in High Energy Physics}
\subsubsection{Conformal Gravity in High Energy Physics}

It has been mentioned above that the fundamental interactions (not considering gravity) are conformally invariant. While apparent at the Lagrangian level, the presence of non-zero masses generally implies that conformal invariance is not a good symmetry in physics. However, considering that masses are now thought to result from coupling to the (now discovered) Higgs boson, a statement that we noted in the literature \cite{9}, indicates that the Standard Model is 'nearly conformally invariant', with the invariance broken by the presence of the mass $m$ of the Higgs field $\phi$ in the scalar Lagrangian
\begin{equation}\label{higgs}
  {\cal L} = \frac{1}{2}\phi_{\mu}^2 - V(\phi), ~~~~V(\phi)= -\frac{1}{2}m^2\phi^2 + \lambda \phi^4
\end{equation}
Conformal invariance can be restored in this Lagrangian by simply replacing the mass term, $(1/2)m^2\phi^2$, with a conformal coupling of the scalar field, namely with the Lagrangian
\begin{equation}\label{higgs-conf}
  {\cal L} = \frac{1}{2}\phi_{\mu}^2 - V(\phi), ~~~~V(\phi)= -\frac{1}{6}R_{\alpha}^{\alpha}\phi^2 + \lambda \phi^4
\end{equation}
However, given the measured value of the mass of the Higgs, $m \simeq 125$ GeV, one can strenuously object to such an approach, considering that the value of the local curvature, computed using the Einstein equations, is much smaller than the square of the Higgs mass $m^2$
\begin{equation}\label{higgs-einstein}
R_{\alpha}^{\alpha} \simeq G_{\alpha}^{\alpha} \simeq \frac{8 \pi}{3}G \rho \simeq \left(\frac{m}{M_{\rm Pl}}\right)^2 m^2 \ll m^2
\end{equation}
where $M_{\rm Pl} \sim 1/\sqrt{G}$ is the Planck mass. On the other hand, if the Ricci scalar $R_{\alpha}^{\alpha}$ is computed within the confines of a purely 4th order theory, like conformal gravity, one can write approximately
\begin{equation}\label{higgs-weyl}
  W_{\alpha}^{\alpha} \simeq (R_{\alpha}^{\alpha})^2 \simeq \frac{1}{\alpha} m^4 \rightarrow R_{\alpha}^{\alpha} \simeq \alpha^{-1/2} m^2
\end{equation}
to make such an identification possible, {\rm assuming} $\alpha \sim O(1)$. Should this be the case, it would imply the potential presence of quantum gravitational effects at the LHC\footnote{There have been suggestions of missing energy at some high energy collisions at LHC, attributed at present to Long Lived Particles (LLP), beyond the Standard Model\cite{10}.}, an issue considered already in the past decade within the framework of additional spatial dimensions\cite{11}.

Actually, considering the central role that the equality of inertial and gravitational masses play in GR, it is strange that the mass, $m$, appearing in the scalar field Lagrangian, associated with spontaneous symmetry breaking and excited at the LHC,  would be so  different from the Planck mass $M^2_{\rm Pl} \simeq 1/G$ that occurs in the gravitational Lagrangian. The disparity of these two masses is known as the \textit{Hierarchy Problem} in high energy physics\cite{9}. Compounding this issue is the fact that loop diagrams of the scalar field appear to diverge quadratically with the upper energy cut-off. An attempt to bring these two mass scales together was proposed in \cite{11}, at the expense of introducing, hitherto unseen, extra space dimensions, while \cite{12} employed these ideas to explain features of the cosmic ray spectrum at the appropriate energies (similar to those considered in \cite{10}).  The point suggested in the present note and in the above discussion, is that, in a yet not fully understood  way,  these two mass scales have a common origin in the same dimensionless gravitational constant but  manifest themselves differently at different spatial or energy scales.

%\smallskip
%\noindent \textit{2. Conformal Gravity and Black Holes}
\subsubsection{Conformal Gravity and Black Holes}

One of the fundamental properties of conformal transformations, the equivalent to a choice of gauge in conformal gravity, is the preservation of the light cone structure. Under this condition, with conformal gravity the underlying theory of gravity, should the Universe begin its expansion without the presence of black holes, it should never be able to form one. One might object to such a statement on the basis of Eq. (2.4) which appears to have at least one horizon near $r \simeq \beta(2-3 \beta \gamma)$. However, this expression represents only a vacuum solution. In GR horizons are possible because,  given the scale of Newton's constant $G$, one can form a horizon by piling up a sufficiently large amount of $M$. It is well understood that for sufficiently large $M$ the tidal forces on the horizon are insufficient to disrupt even a star, let alone gas particles that crosse the horizon, impervious to its presence,  to eventually hit the central singularity.

On the contrary, within the conformal theory, not only the gravitational field, but also (the by necessity conformally coupled) matter are aware of the presence of horizons. We conjecture hence that, in order for the horizon not to be "breached" by in-falling particles, the broken symmetry that provides their mass will be restored rendering them nearly massless; at the same time, what is currently  thought to be the black hole interior, will be replete with the unbroken Higgs vacuum, in a different version of structures called "gravastars"\cite{13}, considered as alternatives to black holes. However, contrary to gravastars,  in the structures implied by these considerations, the vacuum energy that fills their interior would be provided by the restoration of the Higgs symmetry breaking, imposed by the conformal symmetry, in combination with an altogether different effective gravitational constant and likely also quantum gravitational effects.

These considerations are admittedly speculative, but no more so than those of e.g. the Einstein-Rosen bridge or similar ideas; however, they are driven by a novel consideration of the nature of gravitation that appears to address issues  broader than those of black holes. Should these considerations be correct, the implied absence of a horizon  would invalidate the evaporation of black holes and would thus obviate the associated issue of "information paradox" that appears to indicate violation of unitarity in black hole evaporation. As such, the horizons of the known astrophysical black holes, ought to be only apparent horizons (surfaces of huge but finite redshift), with these structures (the known astrophysical black holes) being eternal depositories of (unusable, inaccessible) information, the detritus of gravitational thermodynamic evolution.

\subsection{Astrophysical Considerations}

%\smallskip
%\noindent \textit{1. In Search of G}
\subsubsection{In Search of G}

In our so far analysis we have presented  the reasons motivating a conformal theory of gravity, its exact solutions  and its potential resolution of the galactic rotation curves issues\footnote{For a more extensive discussion of these and different issues of Weyl gravity the interested reader is advised to consult\cite{14}}. The theory is 4th order and because of that it has a dimensionless coupling constant in its Lagrangian. On the other hand, standard Newtonian theory (and its covariant generalization GR) work admirably within most astrophysical domains. Poisson's equation, with its dimensionful coupling constant fares quite well in connecting the gravitational potential to its sources; one would, hence, think this as good evidence for the presence of a constant $G$ value; on the other hand, considering, for example the issue of galactic dark matter, one might consider a constant value of $G$ of only limited validity depending on the scale of the problem. 

In search of qualitative considerations of this issue within Weyl gravity, from Eq. (2.3), by rewriting the action form as
\begin{equation}\label{rmn2}
 I_W = - 2\alpha \int d^4x (-g)^{1/2}[(R_{\lambda\mu}R^{\lambda
\mu}/R^{\alpha}_{\phantom{\alpha} \alpha}-R^{\alpha}_{\phantom{\alpha} \alpha}/3)] R^{\alpha}_{\phantom{\alpha} \alpha}
\end{equation}
one could heuristically consider this form akin to the Hilbert-Einstein action with some average value of the quantity in square brackets playing the role of an average value of $1/G \simeq M^2_{\rm Pl}$, which can now be computed in the exact solution of the full theory. On setting $k = 0$, or for $r^2 \ll 1/k$, this quantity is found to be proportional to $\gamma^2(r^2 - 3 \beta^2)$ arguing in a very qualitative fashion that close to what one consider a black hole horizon the effective Planck mass approaches zero and quantum gravitational effects may become important, or even that gravity might become repulsive if it ever were that $r < \sqrt{3} \beta$, in support of the qualitative arguments given in the previous section.

Should one accept a theory with a dimensionless coupling constant (not necessarily Weyl gravity) as a legitimate gravitational theory, one will still have to invent an effective $G$, since our to date thinking of gravitational dynamics has always been in terms of $G$. This has been the approach of \cite{15} who, motivated by the details of galactic rotation curves have considered them as a result of change in $G$, instead of a change in Newton's $F = ma$ law, as postulated in \cite{2}. However, interpretation of observations without the introduction of unobserved matter, demands introduction of a variation in $G$ at a particular physical scale; while originally considering that to be the acceleration introduced in \cite{2}, they considered as a serious alternative the surface density of the matter involved in the local dynamics, obtaining the following expression for $G$
\begin{equation}\label{DCDK}
  G(s) = G_0 \left[1 + \frac{2}{s + \sqrt{s^2+4s}}\right].
\end{equation}
In this expression $s = \sigma/\sigma_0$, and $\sigma_0$ is a characteristic surface density of order of that given in Eq.(2.6), which, in terms of Newtonian concepts is given by $\sigma_0 = a_0/G_0$. The analysis of \cite{15} includes characteristic values for $G_0$ and $\sigma_0$, however, it indicates how the corresponding dynamics depend on the value of $s = \sigma/\sigma_0$. For a mass restricted in a small region of space, one expects deviation of Newtonian dynamics at $\sigma < \sigma_0$ or $r^2 \simeq (G_0/a_0) M$ or $r_c\sim M^{1/2}$.

\begin{figure}[!h]
\centering\includegraphics[width=2.8in]{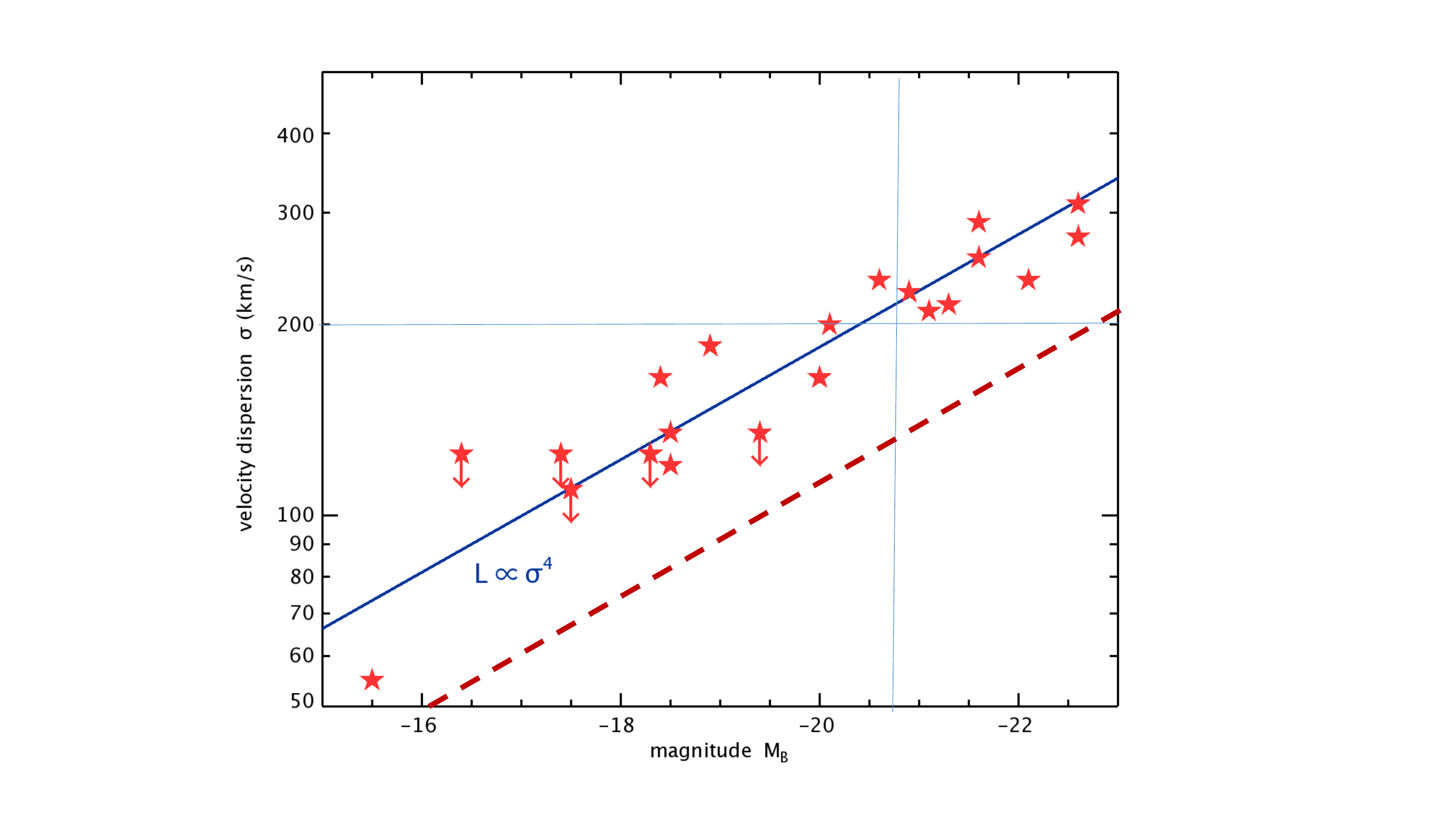}
\includegraphics[width=2.4in]{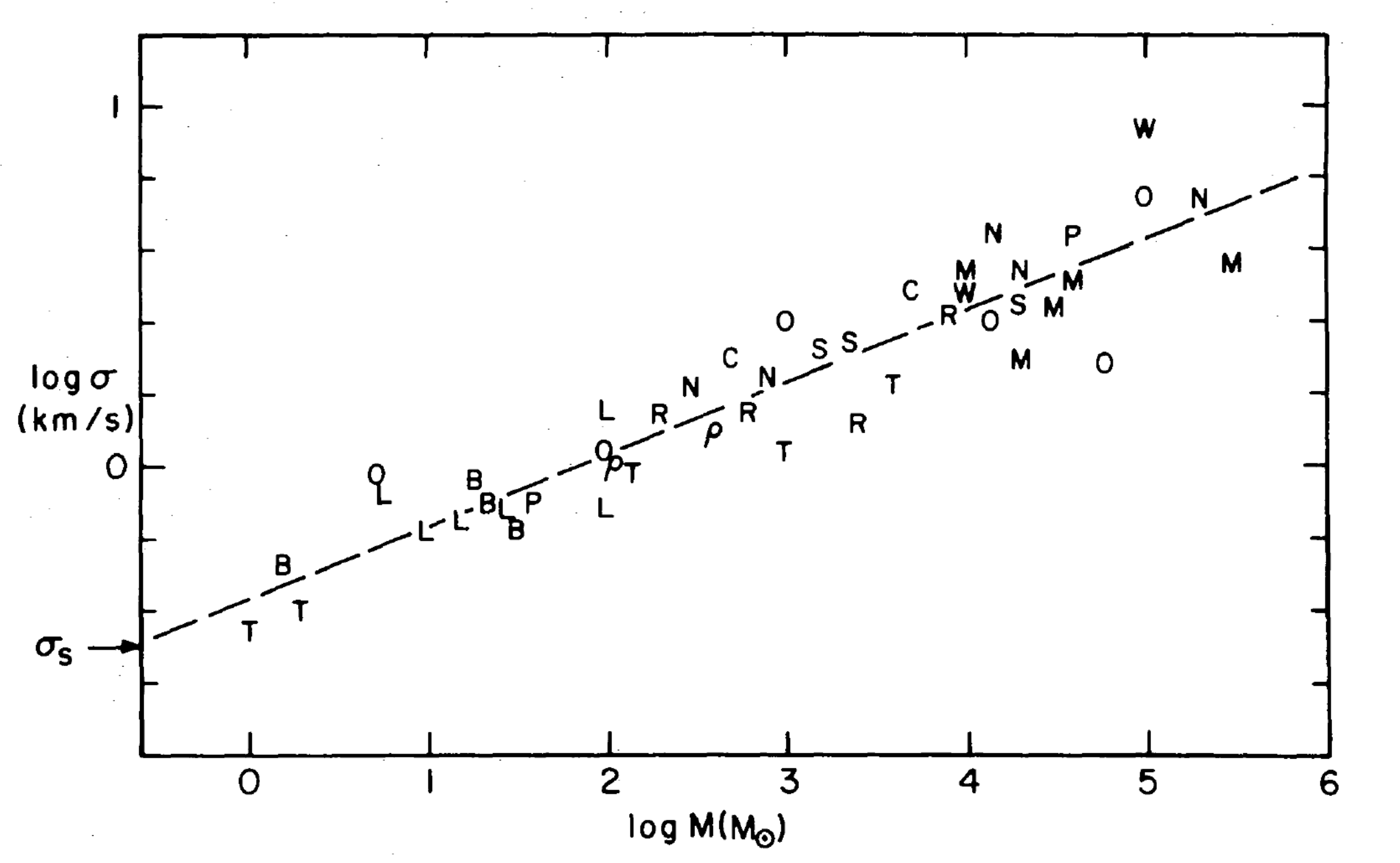}
%%% where xxxxxx name represents "figurename.eps"
\caption{(a) The Faber-Jackson relation of elliptical galaxies, relating their luminosity to their velocity dispersion $\sigma$ (solid blue line; from Wikipedia). The dashed red line is the corresponding mass - velocity dispersion relation obtained from the $L \propto \sigma^4$ relation by the employing the proper mass-to-light ratio of $\simeq 6$ (compared its solar value; absolute magnitude of $M_B \simeq -21$ corresponds to $L \simeq 2 \times 10^{10} L_{\odot}$). (b) The mass - velocity dispersion of molecular clouds  $M \simeq 10^2 (v/1 \,{\rm km\; s}^{-1})^4 M_{\odot}$ (from reference \cite{16}). It should be noted that the dashed line extrapolates into the red dashed line of (a).   }
\label{fig2}
\end{figure}

Neither of the proposals of \cite{2,15} provides detailed dynamics for the formation of bound structures; however should these have a size-mass relation determined by the value of $\sigma_0$, it would lead to a characteristic size, $r_c$, related to their mass by the relation $r_c \sim M^{1/2}$. Assuming further they are in virial equilibrium, one can eliminate the radius between the virial ($ v^2 \simeq M/r_c$) and this size-mass relation to obtain $M \propto v^4$ or more precisely, on substituting the numerical values,
\begin{equation}\label{TFFJ}
M \simeq 10^{11} \left(\frac{v}{200 \; {\rm km/s}}\right)^4 M_{\odot} ~~({\rm solar~ masses}).
\end{equation} 
This relation represents the mass - velocity dispersion relation of elliptical galaxies (dashed line in  figure \ref{fig2}a); this is obtained from their luminosity - velocity dispersion relation (the so-called Faber-Jackson relation -  figure \ref{fig2}a solid line) after correction for the galaxy mass-to-light ratio normalized to the solar values $M/L \simeq 6 \times M_{\odot}/L_{\odot}$.

It is of interest to note that similar relations occur in other astrophysical settings. One of them, associated with galactic molecular clouds, has been known for some time \cite{16} as the Larson Relation (figure \ref{fig2}b). A little known (or perhaps tacitly ignored) fact is that this latter relation (dashed line of figure \ref{fig2}b)  extrapolates into the former Faber-Jackson the relation (dashed line of figure \ref{fig2}a), signifying an underlying common physical reason. As such, we note, in agreement with arguments above, that all these classes of objects have approximately the same column density, which is also the column density of the universe and incidentally that of the electron! These regularities, which span many orders of magnitude in scale, imply a common origin of these facts; while this may  or may not be attributed to Weyl gravity, the presence of the characteristic column density given in Eq.(2.6), obtained first in the exact solution of this theory, does not appear accidental.

%\smallskip
\subsubsection{Gravitational Lensing}
%\noindent \textit{2. Gravitational Lensing}

The exact metric form of Eq. (2.4) incites the computation of gravitational lensing in this theory, as a means of testing other aspects of its astrophysical ramifications. Here, the reader is cautioned that naive application of standard lensing formulae produce nonsensical results (deviation angle increasing with the light ray impact parameter). The reason for such a behavior is due to the fact that standard gravitational lensing formalism has been developed in asymptotically flat spaces. 

Motivated by the approach of \cite{17}, who studied the gravitational lensing effects of the de Sitter component of the Schwarzschild - de Sitter metric, \cite{18} showed that the nonsensical results obtained by applying superficially the asymptotically flat space lensing formalism are due to the difference in the definition of the light bending angles in the non-asymptotically flat space of Eq. (2.4). The result of this approach to the asymptotically non-flat space of Eq. (2.4), for a light ray of impact parameter $b$ with respect to the coordinate origin is given by
\begin{equation}\label{light}
  2 \phi = \frac{4 \beta}{b} - \frac{2 \beta^2 \gamma}{b} - \frac{kb^3}{2 \beta}
\end{equation}
with $\beta, \gamma, k$ the parameters of the vacuum solution of Eq. (2.4). This expression contains the effects of the de Sitter geometry discussed in \cite{17} and the term $2 \beta^2 \gamma/b$ which includes the effects of the $\gamma r$ term of the metric. 

One should note that despite the fact that the $\gamma, k$ terms are associated with conformally flat degrees of freedom and as such they should not contribute to light bending, nevertheless  they do so through their coupling to the non-conformally flat component $\beta$. It is also of interest to note that the dependence of the $\gamma r$ term effect has the same $b$-dependence as the Newtonian $1/r$ term, in distinction to that of the $k$ term, arguing for a different character of these two terms and consistent with the considerations given in \S 2a that one, $\gamma$,  is associated with an interior moment of the source, while the other, $k$, with an exterior one.

\section{Discussion}

We have outlined above the vestiges of a conformally invariant theory of gravity and its exact vacuum and electrovac solutions \cite{3,7,8} and argued that gravity too may need to be recast in terms of a theory relying on a local invariance principle (in reference \cite{14} the interested reader can find a far more detailed discussion of this theory). The point which makes this theory of particular interest is the presence of the linear term in the vacuum solution, which appears to have the correct magnitude of providing stellar dynamics in the outskirts of galaxies that obviate the need of the presence of dark matter there. The important aspect of this circumstance (should that be only a circumstance) is that no such considerations were taken into account in writing down the Lagrangian of the theory (the Lagrangian of Eq. (2.3) has been considered in the past, with the relevant equations for the gravitational field known as the Bach equations; however, in the absence of exact solutions and their observational underpinnings it was not explored in greater detail).

From the purely theoretical point of view, a conformally invariant theory is also of interest because, preserving the light cone structure, it would prevent light cone "tipping" to form gravitational horizons and their enclosed singularities, as is the case with the second order theories.  We have argued qualitatively that this could be possible because the matter that agglomerates to form a black hole would also consist of conformally coupled fields that "do know" about the horizon presence and modify their structure to  prevent the formation of a horizon (i.e. reverse the spontaneous broken symmetry that produced their masses, in combination with quantum gravitational effects that take place in the horizon vicinity). These considerations then open the issue of disparity between the values of the Planck and Higgs scales, each of which provides a mass scale to Physics. While we do not provide answers to this issue in the present note, we have argued that both these disparate masses may have a common origin to a 4th order theory with their different values depending on the scale of the problem.

In relation to this last issue, while not proposing any direct connection to the 4th order conformally invariant theory, we discussed an approach of dynamics with an effective value of $G$ which is not universally constant, but one whose effects depend on the column density of the problem at hand; we argued that such an approach provides also a resolution to the galactic dark matter problem, in regions where the effective column density becomes less than that of the universe, thereby providing a connection to the metric coefficient $\gamma$ of Eq. (2.4).  At the same time, it was argued (in agreement with similar arguments put forward in \cite{2}) that it is this characteristic column that underlies the well known Tully-Fisher and Faber-Jackson relations between the galactic luminosities (or masses) and the fourth power of their stellar velocities, i.e. $M \propto v^4$. It was argued that this same relation extrapolates into the much smaller scales of molecular clouds, implying the potential effects of this  particular column density there too.

In summary, while we do not present a concrete theory that accounts in full detail for the outstanding astrophysical issues of the day, namely the dark matter and dark energy problems, we have shown how the vacuum solutions of a conformally invariant 4th order theory make an immediate contact with these issues. The conformal invariance, the absence of $G$ and the vacuum aspect of the solutions of the theory still need to be integrated into a coherent scheme that can address in detail astrophysical issues. On the other hand, the same principle indicates at the same time a  potential way around some of the issues that have vexed GR and generally 2nd order gravitational theories over many years, such as the singularities and the information paradox in black hole evaporation. In answer to the question in the title of this article, we conjecture that the road to potential progress in all these issues may in fact be trough conformal invariance.

%%%%%%%%%% Insert bibliography here %%%%%%%%%%%%%%


\begin{thebibliography}{9}

\bibitem{1} Frieman, JA, Turner, MS, Huterer, D 2008. {Dark Energy and the Accelerating Universe}. \textit{Ann. Rev. Astron. Astroph.} \textbf{46}, 385

%Cambridge, UK: UIT Cambridge. See \href{http://www.withbotheyesopen.com}{http://www.withbotheyesopen.com}.

\bibitem{2}  Milgrom, M. 1983. A modification of the Newtonian dynamics as a possible alternative to the hidden mass hypothesis. \textit{Astrophys. Journ.} \textbf{270}, 365
    %MacKay DJC. 2008  \textit{Sustainable energy:  without the hot air}.
 %Cambridge, UK: UIT Cambridge. See \href{http://www.withouthotair.com}{http://www.withouthotair.com}.

\bibitem{3} Mannheim, PD, Kazanas, D. 1989. Exact Vacuum Solution to Conformal Weyl Gravity and Galactic Rotation Curves.\textit{Astrophys. Journ.}, \textbf{342}, 635 %Gallman PG. 2011  \textit{Green alternatives and national energy strategy: the facts  behind the headlines}.  Baltimore,\ MD: Johns Hopkins University Press.

\bibitem{4} Eddington, AS. 1922. \textit{The Mathematical Theory of Relativity} (8th Ed., 1960; Cambridge: Cambridge University Press)
%MacKay DJC. 2013.  Solar energy in the context of energy use, energy transportation, and  energy storage. \textit{Proc. R. Soc. A} \textbf{371}.

\bibitem{5} Kazanas, D. \& Mannheim, PD. 1991. General Structure of the Gravitational Equations of Motion in Conformal Weyl Gravity. \textit{Astroph. J. Suppl}, \textbf{76}, 431

\bibitem{6} Begeman, KG. 1989. HI rotation curves of spiral galaxies. I. NGC 3198. \textit{Astron. Astrophys.}, \textbf{223}, 47

\bibitem{7} Mannheim, PD \& Kazanas, D. 1994. Newtonian Limit of Conformal Gravity and the Lack of Necessity of the Second Order Poisson Equation. \textit{Gen. Rel. and Grav.}, \textbf{26}, 337

\bibitem{8} Mannheim, PD \& Kazanas, D. 1991. Solutions to the Reissner-Nordstr\"om, Kerr, and Kerr-Newman problems in fourth-order conformal Weyl gravity. \textit{Phys Rev D}, \textbf{44}, 417
    
\bibitem{9} Meissner, KA. \& Nicolai, H. 2007. Conformal Symmetry and the Standard Model. \textit{Phys. Lett. B}, \textbf{648}, 312
    
\bibitem{10} LHC LLP Community White Paper: Searching for long-lived particles beyond the Standard Modelat the Large Hadron Collider. arXiv:1903.04497
    
\bibitem{11} Arkani-Hamed, N., Dimopoulos, S. \& Dvali, G. 1999. The hierarchy problem and new dimensions at a millimeter. \textit{Phys. Lett. B}, \textbf{429}, 263
    
\bibitem{12} Kazanas, D. \& Nicolaidis, A. 2003. Cosmic Rays and Large Extra Dimensions. \textit{Gen. Rel. Grav.}, \textbf{35}, 1117
    
\bibitem{13} Mazur, PO. \& Mottola, E. 2001. Gravitational Condensate Stars: An Alternative to Black Holes. arXiv:gr-qc/0109035
    
\bibitem{14}. Mannheim, PD. 2006. Alternatives to dark matter and dark energy. \textit{Progr. Part. Nucl. Phys}, \textbf{56}, 340
    
\bibitem{15} Christodolou, D. \& Kazanas, D. Gauss’s law and the source for Poisson’s equation in modified gravity with Varying G. \textit{Mon. Not. R. astr. Soc.}, \textbf{484}, 14
    
\bibitem{16} Larson, RB. 1981. Turbulence and Star Formation in Molecular Clouds. \textit{Mon. Not. R. astr. Soc.} \textbf{194}, 809
    
\bibitem{17} Rindler, W. \& Ishak, M. 2007. Contribution of the cosmological constant to the relativistic bending of light revisited. \textit{Phys. Rev. D}, \textbf{76}, 043006

\bibitem{18} Sultana, J. \& Kazanas, D. 2010. Bending of light in conformal Weyl gravity. \textit{Phys. Rev. D}, \textbf{81}, 127502
\end{thebibliography}
\end{document}